\documentclass[pre,superscriptaddress,showpacs,nofootinbib,floatfix,preprint]{revtex4}
\usepackage{graphicx}
\usepackage{dcolumn}
\usepackage{bm}
\usepackage{epsfig}
\begin{document}

\preprint{PREPRINT}

\title{Thermodynamic and Dynamic Anomalies for Dumbbell Molecules
Interacting with a  Repulsive Ramp-Like Potential}

\author{Paulo A. Netz}
\affiliation{Instituto de Qu\'{\i}mica, Universidade
Federal do Rio Grande do Sul,  Porto Alegre, RS, BRAZIL}

\author{Sergey V. Buldyrev}

\affiliation{Yeshiva University, Department of Physics, 500 West 185th 
Street, New York, NY 10033, USA}
\affiliation{Center of Polymer Studies and Department
of Physics, Boston University Boston, MA 02215, USA}

\author{Marcia C. Barbosa}
\affiliation{ Universidade Federal do Rio Grande do Sul, Caixa Postal
15051, 91501-970, Porto Alegre, RS, BRAZIL} 
\email{marcia.barbosa@ufrgs.br}

\author{H. Eugene  Stanley}
\affiliation{Center of Polymer Studies and Department
of Physics, Boston University Boston, MA 02215, USA}

\date{EH18077-25 March -- dumbbell.tex}

\begin{abstract}

Using collision driven discrete molecular dynamics (DMD), we 
investigate
the thermodynamics and dynamics of systems of 500  dumbbell molecules
interacting by a purely repulsive ramp-like discretized potential,
consisting of $n$ steps of equal size. We compare the behavior of the two
systems, with $n = 18$ and $n = 144$ steps.  Each system exhibits both
thermodynamic and dynamic anomalies,  a
density maximum and the translational and rotational mobilities 
show anomalous behavior. Starting with very dense systems and 
decreasing
the density, both mobilities first increase, reache a maximum, then
decrease, reache a minimum, and finally increase;  this behavior is
similar to the behavior of SPC/E water. The regions in the
pressure-temperature plane of translational and rotational mobility
anomalies depend strongly on $n$. The product of the translational
diffusion coefficient and the orientational correlation time 
increases with  temperature, in contrast with the
behavior of most liquids.

\end{abstract}

\maketitle

\section{Introduction}

Recently much attention has focused on phase behavior of single
component systems comprised of particles interacting via core-softened
potentials
\cite{pabloreview,Ja98,St98,Sc00,Fr01,Bul02,fran02,Wi02,Camp05}.
Core-softened potentials exhibit a repulsive hard core plus a softening
region, which can be a linear or nonlinear repulsive ramp, a shoulder,
a double (or multiple) attractive well, or  combinations of all these
features. These models were created with the purpose of constructing a
simple two-body isotropic potential capable of describing some aspects
of the anomalous behavior of complex fluids \cite{Fr85,Ne01,Gr04}, like
maximum in density as a function of temperature, increasing 
isothermal compressibility upon cooling, and increasing  
molecular
diffusivity with the increase  of pressure \cite{Ne01}.  It has been
proposed some time ago that these anomalies might be associated with a
critical point at the terminus of a liquid-liquid line, in the unstable
supercooled liquid region \cite{Po92}.  The study of core-softened
potentials generates models that are computationally (sometimes even
analytically) tractable and that may retain some qualitative features 
of network forming  fluids such as water.

Hemmer and Stell \cite{hemmer70}, using the method of Takahashi
\cite{takahashi66} proposed the possibility of a second critical point
in addition to the normal liquid-gas critical point, for a
one-dimensional system whose pairwise potential possesses a region of
negative curvature with respect to separation distance in its repulsive
core.  Debenedetti et al. developed thermodynamic arguments to show 
that
this type of potential can lead to density anomaly \cite{De91}.
Stillinger and Stillinger \cite{St97} showed that a pure repulsive
Gaussian core-softened potential produces both thermodynamic and 
dynamic
anomalies but no indication of the second critical point at positive
temperatures.

A number of works investigated step potentials consisting of a hard
core, a square repulsive shoulder and an attractive square well
\cite{St98,Sc00, Fr01,Bul02,Bu02,Bu03,Sk04,cho96,cho96b,He05,Ol06}. In two
dimensions, such potentials have density and diffusion anomalies, a
negatively sloped freezing line and possibly a second critical point in
a deeply supercooled state inaccessible by simulations. In three
dimensions, these potentials do not have dynamic and thermodynamic
anomalies but possess a second \cite{Sk04} and sometimes a third
\cite{Bu03} critical point, accessible by simulations in the region
predicted by the hypernetted chain integral equation \cite{Fr01,Ma04}.

Jagla proposed a different version of the core-softened potential, the
main element of which is a linear repulsive ramp \cite{Ja98}. These
potentials show a region of density anomaly in the pressure-temperature
plane and also display a liquid-liquid critical point, which can be
located in a stable or metastable fluid region, depending on the choice
of parameters \cite{gibson-cond-mat-2006}.  The Rogers-Young 
\cite{Ro84}
approximation, together with the mode coupling theory, reproduces these
features analytically \cite{Ku04}. An attractive ramp \cite{Ja98,Wi02,Xu05}
displays a normal gas-liquid critical point and also brings this second
critical point into an accessible region of higher temperature.
   
An interesting issue, besides the thermodynamic anomalies, is the
investigation of the dynamic anomalies, such as those exhibited by
water.  The translational diffusion coefficient $D$ for a spherical
particle solute of radius $\sigma$ in a medium of shear viscosity
$\eta$, is predicted by the Stokes-Einstein relation to be
\begin{equation}
\label{D}
 D=\frac{k_BT}{6\pi\eta \sigma}\; .
\end{equation}
The rotational autocorrelation time is predicted by the
Debye-Stokes-Einstein relation to be
\begin{equation}
\label{tau}
\tau_r=\frac{4\pi\eta \sigma^3}{3k_B T}\; .
\end{equation}

Simulations of supercooled SPC/E water at constant temperature show 
that
$D$ increases as pressure $P$ decreases, reaches a maximum at a density
$\rho_{D{\rm max}}$ and decreases until $D$ reaches a minimum at
$\rho_{D{\rm min}}$ \cite{Ne01,Ne02a,Ne02b,Ne02}.  Further $\tau_r$, 
has
a minimum at $\rho_{D{\rm max}}$ and a maximum at $\rho_{D{\rm min}}$
\cite{Ne01,Ne02a,Ne02b,Ne02}, and the product $\tau_r D$ is almost
constant with density and temperature \cite{Ne02a,Ne02b,Ne02}. This
constancy is still an open issue and needs to be resolved.
 
Combining Eq.~(\ref{tau}) with Eq.~(\ref{D}), we see that the product
$\tau_r D$ is constant, which could explain the constancy of the 
product
$\tau_r D$ observed the SPC/E water.  However, the hydrodynamic
approximations used in both the Stokes-Einstein and the
Debye-Stokes-Einstein relations are valid only if the size ratio 
between
the solute and the solvent is large \cite{Ka00}.  This is not the case
of the self diffusion of SPC/E water where the tracer particle has the
same size as the other particles in the system, so hydrodynamic
arguments cannot explain the constancy of $\tau_r D$ observed in water
\cite{Ne02a,Ne02b,Ne02}.

Furthermore, for a number of supercooled liquids, experimental results
show that there is a breakdown of the Stokes-Einstein relation close to
the glass transition temperature $T_g$ \cite{Ng99,Fu92,Ci95}. The
translational diffusion close to $T_g$ behaves as $D\propto T/\eta^b$
with $b<1$ \cite{Ng99}.  In this case, the product $\tau_r D\propto
\eta^{1-b}$ is not constant, but increases as the system is cooled
\cite{Ng99}. Since water differs from other supercooled liquids by the
presence of thermodynamic anomalies, the study of the relation between
the thermodynamic and dynamic anomalies in water might shed some light
on this problem.  For SPC/E water, the region in the $P-T$ phase 
diagram
where water has diffusivity anomalies defines a broad region, and the
region where density anomalies can be found lies entirely inside the
anomalous diffusivity region 
\cite{Ne01,Ne02a,Ne02b,Ne02,errington2001}. For SPC/E water, the product
$\tau_r D$ is approximately constant inside the entire region of the 
diffusivity anomaly. 

The goal of our study is to test the hypothesis
that the constancy of $\tau_r D$ is the consequence of the diffusivity anomaly. To 
achieve
this goal we construct a simple model which has the thermodynamic and
dynamic anomalies of water but also has rotational degrees of freedom,
so that  $\tau_r$ can be investigated. We based our model on
the Jagla repulsive ramp model which has the entire spectrum of water 
anomalies~\cite{Ku04}, including the correlation between the 
translational and orientation order parameters \cite{Yan}. In order 
to create rotational
degrees of freedom, we linked two ramp particles into a ``dumbbell'' by a 
permanent bond equal in length to the hard core diameter. Thus, in our
model the two monomers of the dumbbell touch each other with their hard-cores
as in rombohedral crystal obtained in \cite{Ku04} near the high 
pressure boarder of the diffusivity anomaly region.
We keep the same choice of the ratio of the soft-core to the hard-core 
as in~\cite{Ku04}, since these ratio reproduces also the water-like 
correlation of the translational and orientational order maps  \cite{Yan06}.
Since hard-core models cannot be easily treated with a continuous molecular
dynamics, we discretized the ramp by a step-function consisting of many
small steps and use the discrete molecular dynamic (DMD) algorithm  
as in ~\cite{Ku04}.  

First, we  test whether the thermodynamic and dynamic anomalies are
preserved for dumbbells and how their region in the $P-T$ diagram is
alternated comparatively to the original monomeric particles. 
Since it ~\cite{Ku04} it was shown that the dynamic properties 
depend significantly on the step of the discretized potential,
we test if our conclusions are not an artifact of the discretization.
Finally, in the density anomaly region, we can test if the
product $\tau_r D$ is constant (as in water) or if it resembles the
behavior of other supercooled liquids. 

This paper is organized as follows. We introduce the dumbbell model and
give details about the simulations in Sec.~II. We present the
thermodynamic and dynamic behavior in Sec.~III, and our conclusions in
Sec.~IV.

\section{The Dumbbell  Model}

The model we study is defined as follows. We consider a set of diatomic
molecules formed by two spherical atoms of diameter $\sigma_0$ linked
together.  The atoms in each molecule are separated from each other by 
a distance $\ell$ which is allowed to fluctuate in the range
$0.99\sigma_0\leq\ell\leq 1.01\sigma_0$.  The interaction potential
between two atoms belonging to different dumbbell molecules separated 
by
a distance $r$ is a ``ramp'' made of $n$ steps \cite{Ku04}
(Fig.~1). Thus
\begin{equation}
\label{pot}
U (r) \equiv\cases{\infty & $r<\sigma_0$ \cr k\Delta U & 
$r_{k+1}<r<r_k$
                   \cr 0 & $r>\sigma_1-\frac{\Delta r}{2}$} \; ,
\end{equation}
where
\begin{eqnarray}
r_k &\equiv& \sigma_1 - \left(k-{1\over 2}\right)\Delta r\;,\\ \Delta r
&\equiv& {\sigma_1 - \sigma_0\over n+(1/2)}\; ,
\end{eqnarray}
and
\begin{equation}
\Delta U \equiv {U_0\over n+(1/2)} \; .
\end{equation}

The liquid phase part of the ramp phase diagram has the following
characteristics.  At low $P$, particles prefer to be at distances
$\sigma_1$ from each other. At high $P$ the typical distance is
$\sigma_0$.  This implies a crossover region of $P$ with anomalously
large isothermal compressibility $k_T$ which might give rise to an
accessible liquid-liquid phase transition if an attractive part is
included in this potential.

The discretized, step version of the ramp potential has the advantage 
of being well suited for study using collision-driven molecular dynamics
\cite{rapaport95,Sk04}.  As $n\rightarrow \infty$, the step potential
becomes a ``ramp'' similar to the potential introduced by Hemmer and
Stell \cite{hemmer70} and studied by Jagla \cite{Ja98}.  Indeed, the
properties become similar to a pure ramp potential if the height
difference between two adjacent steps is less than $k_B T$
\cite{Ku04}. Moreover, we expect \cite{Ku04} that the dynamic behavior
converges to the behavior of the smooth ramp when $n \to \infty$ as
O($n^{-1}$), therefore in a slower way than the thermodynamic
properties, which converge as O($n^{-2}$).

In order to test the influence of the discretization, we study this
repulsive ramp potential with $n=18$ and $n=144$ steps and
$\sigma_1/\sigma_0$ = 1.74.  We perform ``discrete molecular dynamics''
(DMD) simulations, using a collision-driven algorithm
\cite{fran02,rapaport95,Sk04}, for a system comprised of 500 dumbbell
molecules (1000 ``atoms'') in a cubic box, with box edge ranging from
$L=13\sigma_0$ to $L=15.5\sigma_0$.  The dimensionless temperature
$T^\ast$, density $\rho^\ast$, pressure $P^\ast$, translational
diffusion coefficient $D^\ast$, and mean rotational correlational time
$\tau_r^\ast$ are given respectively by
\begin{equation}
\label{T*}
T^*\equiv \frac{T k_B}{U_0},
\end{equation}
\begin{equation}
\label{rho*}
\rho^*\equiv\rho \sigma_0^3,
\end{equation}
\begin{equation}
\label{P*}
P^*\equiv\frac{P \sigma_0^3}{U_0},
\end{equation}
\begin{equation}
\label{D*} 
D^*\equiv\frac{D(m/U_0)^{1/2}}{\sigma_0},
\end{equation}
and
\begin{equation}
\label{tau*}
\tau_r^*\equiv\frac{\tau_r(U_0/m)^{1/2}}{\sigma_0}\;,
\end{equation}
where $k_B$ is the Boltzmann constant, $m$ the particle mass, and $U_0$
the high of the repulsive ramp.

We calculate thermodynamic and dynamic properties using simulations of
160,000 time steps. We analyze the thermodynamic and dynamic behavior
for dimensionless temperatures ranging from $T^*=0.151$ down to
$T^*=0.054$ for $n=18$ and from $T^*=0.168$ down to $T^*=0.048$ for
$n=144$.

\section{Results}

\subsection{Density Anomaly}

We shall first investigate if our system of diatomic molecules exhibits
a density maximum as found for a system of spherical monoatomic
particles interacting through the same potential \cite{Ku04}.  Figure 2
plots the pressure $P^*$ against density $\rho^*$ for fixed values of
$T^*$. We see that for all the states there is no mechanical
instability, and the pressure is a monotonically increasing function of
density,
\begin{equation}
\label{prhoT}
\left( \frac{\partial P}{\partial \rho} \right)_{T} > 0 \; ,
\end{equation}
implying that no phase transition is present. The isotherms cross each
other, which means that $(\partial \rho/\partial T)_{P}$ = 0, which
implies a density anomaly.

In order to locate the points where $(\partial P/\partial T)_{\rho}$ =
0, we plot $P^*$ against $T^*$ along isochores (Fig.~3a, 3b and 3c), 
fit the
results to a polynomial, and calculate the minima. The line connecting
these minima is the temperature of maximum density (TMD) line, which is
the boundary of a thermodynamically anomalous region where
\begin{equation}
\label{ptrhoanom}
\left(\frac{\partial P}{\partial T} \right)_{\rho} < 0 \; ,
\end{equation}
and therefore an anomalous behavior of the density (similar to water)
\begin{equation}
\label{rhoTp>}
\left( \frac{\partial \rho}{\partial T} \right)_{P} > 0 \; .
\end{equation}
Because of the Maxwell relations, Eq.~(\ref{rhoTp>}) implies anomalous
behavior of the entropy,
\begin{equation}
\label{Srho}
\left( \frac{\partial S}{\partial \rho} \right)_{T} > 0 \; .
\end{equation}

We can uncover the effect of increasing  $n$ on generating the thermodynamic
and dynamic anomalies, comparing Figs.~3a and 3b. We will see that the
number of steps has almost no influence on the shape and location of 
the
TMD, consistent with the fast convergence of pressure to the linear 
ramp
value as $n \to \infty$ \cite{Ku04}. Therefore, the curves for the
system with $n=144$ steps are expected to be almost identical to the
curves for the linear ramp \cite{gibson-cond-mat-2006}.  The
corresponding effect of the dumbbell shape can be estimated comparing
the results of dumbbells with $n=18$ and $n=144$ with the systems of
1000 monoatomic particles with $n=18$ (Fig. 3c) and $n=144$ \cite{Ku04}
interacting with the same potential illustrated in Eq.~(\ref{pot}) for
the same state points as the dumbbell molecule system.  The systems of
interacting dumbbells have a slightly smaller region of density (and
entropy) anomaly, shifted to higher pressures. In these cases, for
densities above 0.182 and below 0.154, no minima in the pressure are
found.
 
\subsection{Dynamics}

We now study the mobility associated with the repulsive ramp $n$-step
potential for $n=18$ and $n=144$. We calculate the translational
diffusion coefficient $D$ using the mean-square displacement averaged
over different initial times,
\begin{equation}
\langle \Delta r^2(t) \rangle = 
\langle [{\bf v}(t_0+t)-{\bf r}(t_0)]^2\rangle\; 
\end{equation}
where ${\bf v}$ is the center of mass of the dumbbell.
Then $D$ is obtained from the relation
\begin{equation}
D=\lim_{t\to\infty}\langle \Delta r^2(t) \rangle/6t \; .
\end{equation}

Figure 4 shows the behavior of the dimensionless translational 
diffusion
coefficient, $D^*$, as a function of the dimensionless density,
$\rho^*$, for the dumbbell potential. At low temperatures, the behavior
is similar to the behavior found for SPC/E supercooled water
\cite{Ne01}. $D$ increases as $\rho$ decreases, reaches a maximum at
$\rho_{D{\rm max}}$ (and $P_{D{\rm max}}$), and decreases until it
reaches a minimum at $\rho_{D{\rm min}}$ (and $P_{D{\rm min}}$). Above
$T^* = 0.086$ (not shown), however, the anomalous behavior disappears.

The region in the $P-T$ plane where there is an anomalous behavior in
the diffusion is bounded by $(T_{D{\rm min}}(\rho),P_{D{\rm
min}}(\rho))$ and $(T_{D{\rm max}}(\rho),P_{D{\rm max}}(\rho))$.  The
location of this region for the studied systems is shown by the dotted
lines in Figs.~3a, 3b, and 3c. The effect of the dumbbell shape 
observed
by comparing Fig.~3a with 3c, or Fig.~3b with Ref.~\cite{Ku04}, is
rather small and affects only the overall shape of the curve, shrinking
slightly the region and shifting toward higher values of P and lower
values of T. The effect of $n$, obtained by comparing Fig.~3a with 3b 
or
Fig.~3c with Ref.~\cite{Ku04}, however, is very strong. Whereas in the
systems where the potential was discretized in 144 steps (Fig.~3b and
Ref.~\cite{Ku04}) the diffusion anomaly line lies outside the TMD line
(exactly as in SPC/E water\cite{errington2001}), for the systems with
$n=18$ (Fig.~3a and 3c) the diffusion anomaly region shrinks strongly,
in such a way that it migrates to a region inside the TMD line, in
contrast to the behavior of SPC/E water.  These results are consistent
with the slow convergence of the dynamic properties of the discretized
ramp potential, as $n \to \infty$. Nevertheless, the results for 
$n=144$
are expected to be very similar to the results for a linear ramp.

The use of dumbbell molecules, unlike spherically symmetric models,
allows us to calculate the rotational degrees of freedom and to 
estimate
of the role of anisotropy. The rotational diffusion is analyzed by
calculating the orientational autocorrelation function of the vector
${\bf e}(t)$ defining the orientation of the dumbbell molecule,
\begin{equation}
\label{Ce}
C_{\bf e} (t) \equiv \langle {\bf e}(t) \cdot {\bf e}(0)\rangle.
\end{equation}
The orientational autocorrelation function depends on the density and
temperature, and for short times we can fit $C_{\bf e}(t)$ with an
exponential function \cite{note2}:
\begin{equation}
\label{expontau}
C_{\bf e} (t) \propto \exp (- t/\tau_r).
\end{equation}
The orientational correlation times, $\tau_r$, obtained from
Eq.~(\ref{expontau}) depends on density in a roughly complementary way
as the translational diffusion coefficient $D$.  This behavior is shown
in Figs.~5a (for the $n=18$ system) and 5b (for the $n=144$ system).

As $\rho$ decreases, $\tau_r$ decreases, reaches a minimum, then
increases, reaching a maximum. A similar behavior was also observed for
SPC/E water simulations \cite{Ne02a,Ne02b,Ne02}. The density region
where the rotational diffusion is anomalous roughly coincides with the
region of translational diffusion anomalies, as seen in Figs.~3a and 3b
(dashed lines).  Here again the anomaly region depends strongly on $n$.

The translational and rotational diffusion for systems for the $n$-step
repulsive ramp seem to behave in a complementary way, but in fact the
product $\tau_rD$, as shown in Fig.~6a, is not density or temperature
independent as in SPC/E water \cite{Ne02a,Ne02b,Ne02}. Indeed, $\tau_r
D$ increases with increasing temperature (Fig.~6b), in sharp contrast
with most liquids.  Inside the thermodynamically anomalous region the
product $\tau_rD$ is also not a constant, and a careful examination of
the very low temperature behavior (Fig.~6b) shows that the product
increases upon cooling below $T^*=0.048$,  in a way similar to 
many other systems close to the glass transition temperature \cite{Ci95}. 
For our systems we estimate the Mode Coupling Temperature 
to be $T_{MCT}^*=0.044\pm 0.001$ and using the Vogel-Fulcher ansatz 
$D\propto  \exp(A/(T-T_0))$ we 
estimate   $T_0^*=0.038\pm 0.001$. The glass transition temperature is
between these two values.

The explanation for the anomalous behavior of the product $\tau_r D$ is
that the rotation of the dumbbell in the vicinity of other dumbbells 
may
have a different activation energy than translation. This is quite
plausible from simple geometrical considerations. In order to rotate,
the dumbbell must overcome a larger potential barrier than to 
translate.
Then $\ln \tau_{r}D= (A_{r} -A_{t})/kT$, where both $A_{r}$ and $A_{t}$
should be functions of density. Indeed the Arrhenius plot of 
$\tau_{r}D$
at high and moderate temperatures gives perfect straight lines with 
weak
dependence on density (Fig.~6b).  Figure 7 shows that $A_r-A_t$, which
quantifies how strongly the product $\tau_r D$ deviates from a 
constant,
reaches its maximum close to the density corresponding to the closed
packed arrangements of the rumps, below which the dumbbells may 
translate without overcoming any potential barrier. 
This density may be estimated for an arrangement of dumbbells
consisting of triangular lattices of vertically placed dumbbells
$\rho=1/(\sigma_1^3/\sqrt{2}+\sigma_1^2 \sigma_0
\sqrt{3}/2)=0.153\sigma_0^{-3},$ where $\sigma_0$ is the hard core and
$\sigma_1$ is considered a soft core diameter. The linearity of the
Arrhenius plot breaks down close to the glass transition where 
$\tau_{r} D$ starts to increase upon cooling as in other liquids in
which this behavior is usually associated with the 
breakdown of the Einstein-Stokes 
caused by the growth of dynamic heterogeneities near the glass transition 
\cite{Ng99,Ci95,Ma05}.

\section{Conclusions}

We have simulated a set of dumbbell molecules interacting through a
purely repulsive ramp-like potential, discretized in $n$ steps.  We
compare the cases of $n=18$ and $n=144$, where the $n=144$ case has
the behavior very similar to that if the
 linear ramp.  We studied the density
anomaly and anomalies in both the translational and rotational
mobilities.  We found that in both cases the density behaves anomalously in a
certain range of pressures and temperatures.  Comparison with
simulations of monoatomic particles interacting with the same potential
\cite{Ku04} shows that the thermodynamic anomalies are not very
sensitive to the number of steps used in the potential, and only weakly
affected by the dumbbell shape. Thus the existence of a fixed link 
between pairs of atoms and the resulting anisotropy has only a small effect on
the thermodynamic properties.

The dynamic anomalies, however, are more sensitive, both to $n$ and to
the introduction of the dumbbell shape.  The translational diffusion
coefficient at constant temperature has a maximum and a minimum as a
function of $\rho$, but the locus of anomalous behavior depends both on
the particles' shape (monoatomic or dumbbell) as well as on the number
of steps (a potential with 144 steps has a different range than a
potential with 18 steps). The dumbbell shape causes the dynamically
anomalous region to shrink. The decrease of $n$ shrinks and shifts the
dynamically anomalous region to lower temperatures.

The region of dynamic anomalies lies approximately inside the 
region of thermodynamic
anomalies for the $n=18$ potential and outside for the $n=144$
potential. We expect that systems with molecules interacting with a
a smooth linear ramp potential would display a behavior similar to
this last one. Despite of the fact that this last hierarchy of anomalous
regions is similar to that found in simulations of water, some 
important differences should be pointed out.  For dumbbell molecules,
the product
of the translational diffusion constant $D$ and the orientational
correlation time $\tau_r$ is not a constant, even at high temperatures,
indicating that translation and rotation of dumbbells may have 
different activation energies. This is not the case in water in which
the molecules are confined in symmetrical tetrahedral shells, so the 
translation
and rotation are associated with switching the partners via bifurcation 
of the hydrogen
bonds \cite{Ne02a}. It would be interesting to compare the behavior 
of the $\tau_r D$ 
in the dumbbell model
and other dimeric liquids, such as hydrogen peroxide. 
At low temperatures, the dumbbell model displays a sharp increase of
$\tau_r D$, resembling the behavior found for liquids close to $T_g$
\cite{Ci95}. This behavior must be investigated by analyzing the
``fastest'' rotational and translational clusters 
\cite{Ki04,Ma05,Jung04}.

\subsection*{Acknowledgments}

We thank P. Kumar for helpful insights and comments on the manuscript,
and the Brazilian science agencies CNPq, FINEP  and Fapergs, and
by the NSF Chemistry Program for financial support. SVB thanks
the office of Academic affairs of the
Yeshiva University  for finantial support.

\newpage
\begin{figure}[p]
\label{figsteps}
\includegraphics[clip,width=12cm]{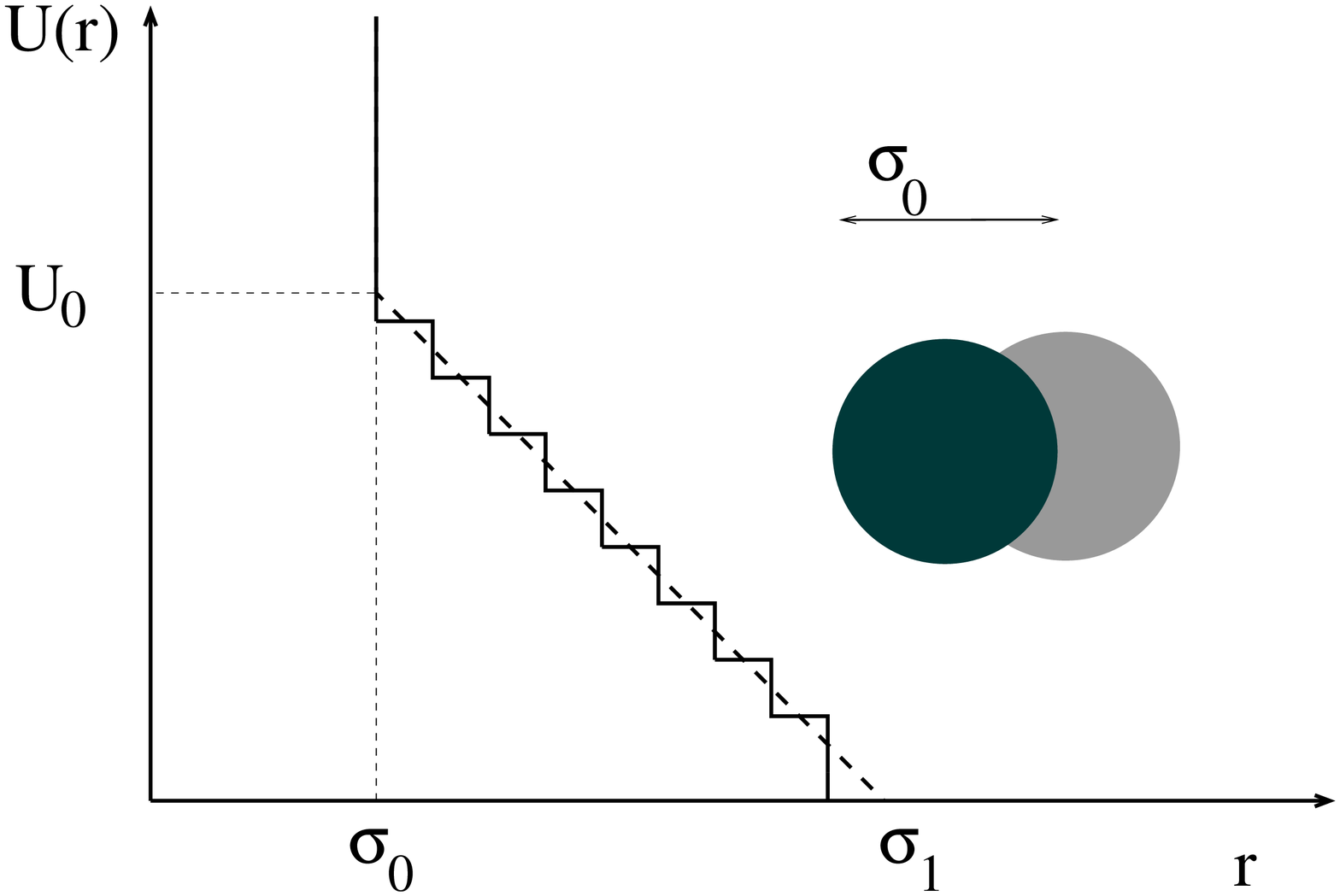}
\caption{Purely repulsive ramp-like discretized potential consisting of
$n$ steps of equal size. Step potential (solid line) and comparison 
with
the linear ramp (dashed line). Here $\sigma_0$ is the diameter of a
spherical atom forming the dumbbell.}
\end{figure}

\newpage

\begin{figure}[p]
\label{fig2}
\includegraphics[width=12cm]{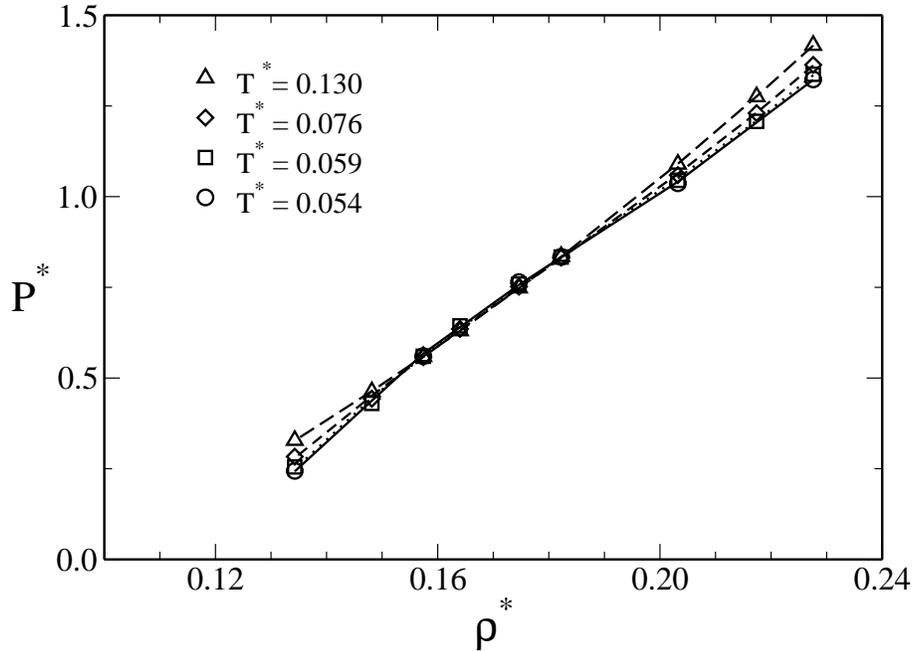}
\caption{Dimensionless pressure, $P^*$, as a function of dimensionless
density, $\rho^*$, for four different isotherms in the system with
$n=18$ steps (Fig.~1). Note that the pressure isotherms cross each 
other
(dashed circle), which implies a density anomaly since $d\rho/dT=0$.}
\end{figure}

\newpage

\begin{figure}
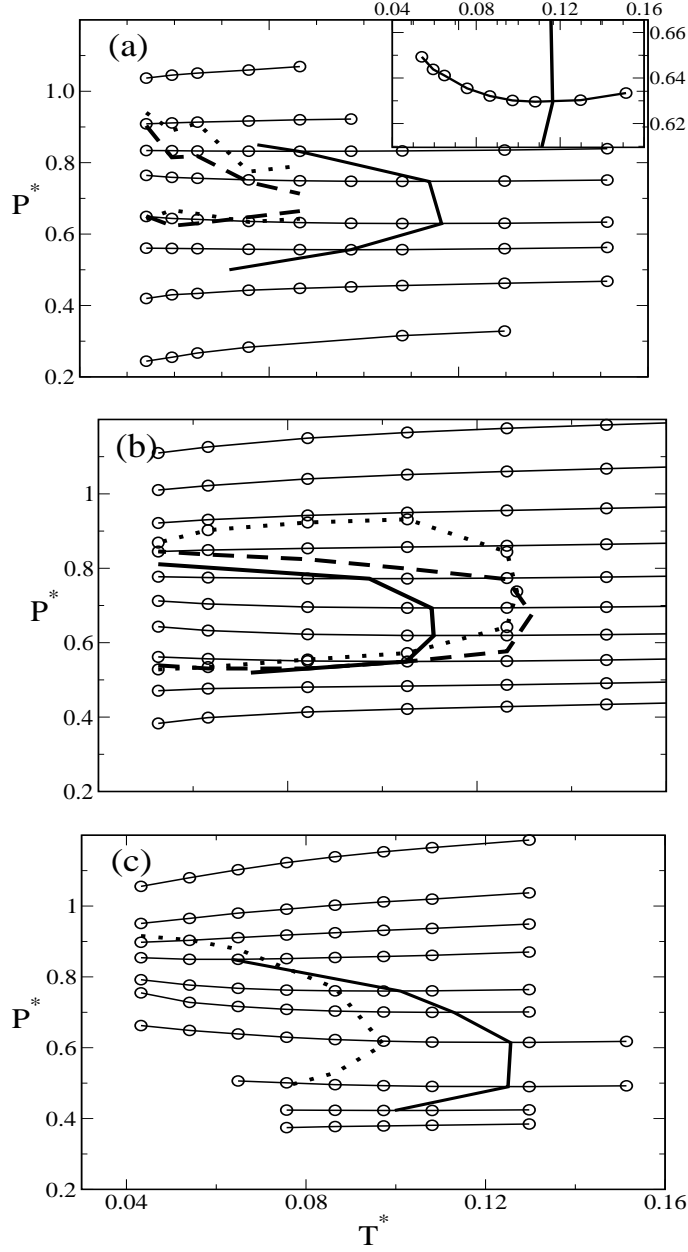

\label{fig3ab}

\includegraphics[clip,width=9cm,height=5.5cm ]{presstemp_dumb_ins.eps}

\includegraphics[clip,width=8.5cm,height=5.5cm]{presstemp_dumb_144.eps}

\includegraphics[clip,width=9cm,height=5.5cm]{presstemp_mono.eps}

\caption{Dimensionless pressure against dimensionless temperature along
isochores. Bold line: TMD line. Dotted line: boundary of the 
diffusivity
extrema.  Dashed line: boundary of the rotational mobility extrema. (a)
Dumbbell molecules interacting via the repulsive discretized ramp $p$
potential with $n=18$ steps. From top to bottom, densities
$\rho^*=0.203$, 0.190, 0.182, 0.175, 0.164, 0.157, 0.148, and 0.134. In
the inset it is shown the density $\rho^*=$ 0.164, in a scale suitable
to see the minimum in the pressure.  (b) The case of $n=144$ steps.
From top to botton, densities $\rho^*=0.208$, 0.199, 0.190, 0.182,
0.175, 0.167, 0.161, 0.154, 0.148, and 0.142.  (c) System of 1000 atoms
(without the bonds), with interaction potential with $n=18$ steps. From
top to bottom, densities $\rho^*=0.406$, 0,381, 0.364, 0.349, 0.328,
0.315, 0.296, 0.269, 0.254 and 0.244. }
\end{figure}

\newpage

\begin{figure}
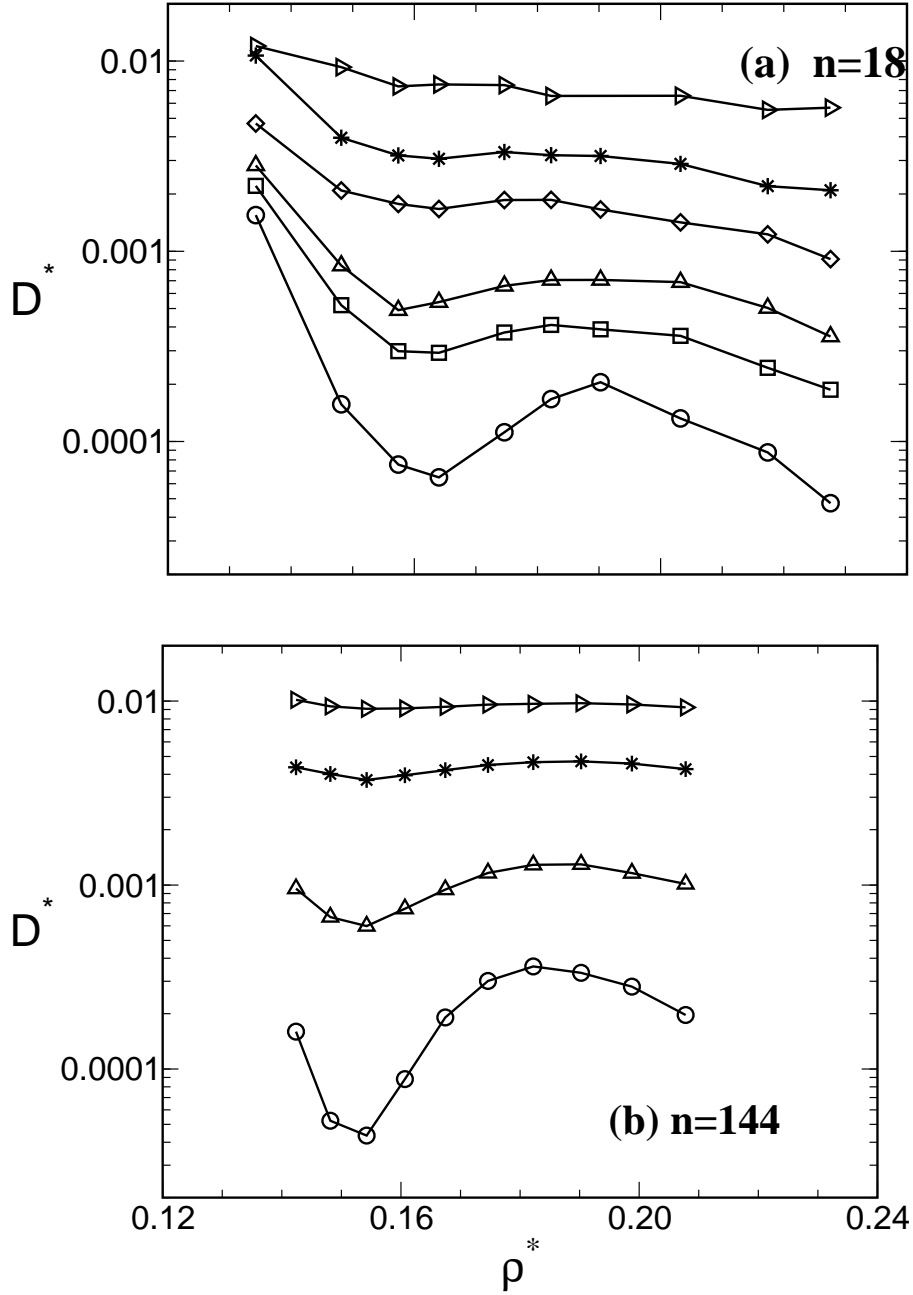

\label{fig4ab}
\includegraphics[clip,width=12cm]{diff_dumb.eps}
\includegraphics[clip,width=12cm]{diff_dumb_144.eps}
\caption{(a) Diffusion coefficient as a function of
density, along six isotherms (from top to bottom
$T^{*} = $ 0.108, 0.086, 0.075, 0.065, 0.059, 0.054)
for $n=18$ steps.
(b) $T^{*} = $ 0.105, 0.084, 0.063, 0.052 
for $n=144$ steps. }
\end{figure}

\newpage

\begin{figure}
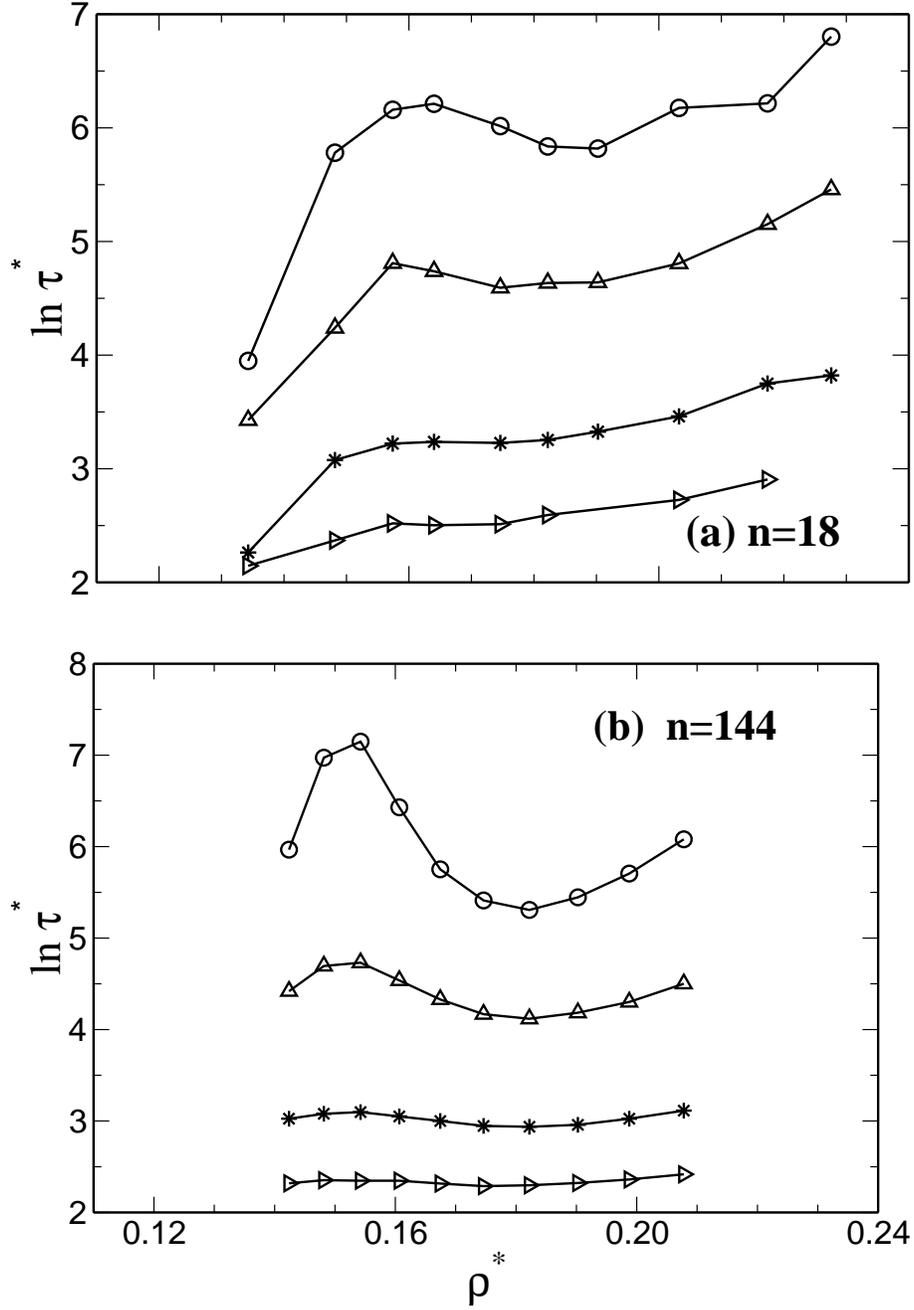

\begin{center}
\label{fig5ab}
\includegraphics[clip,width=12cm]{lntau.eps}
\includegraphics[clip,width=12cm]{lntau-144.eps}
\end{center}

\caption{(a) Orientational time as a function of
density for four isotherms 
(from bottom to top $T^{*} = $ 0.108, 0.086, 0.065, 0.054)
 for $n=18$ steps (b) $T^{*} = $ 0.105, 0.084, 0.063, 0.052
for $n=144$ steps. }

\end{figure}

\newpage


\newpage

\begin{figure}
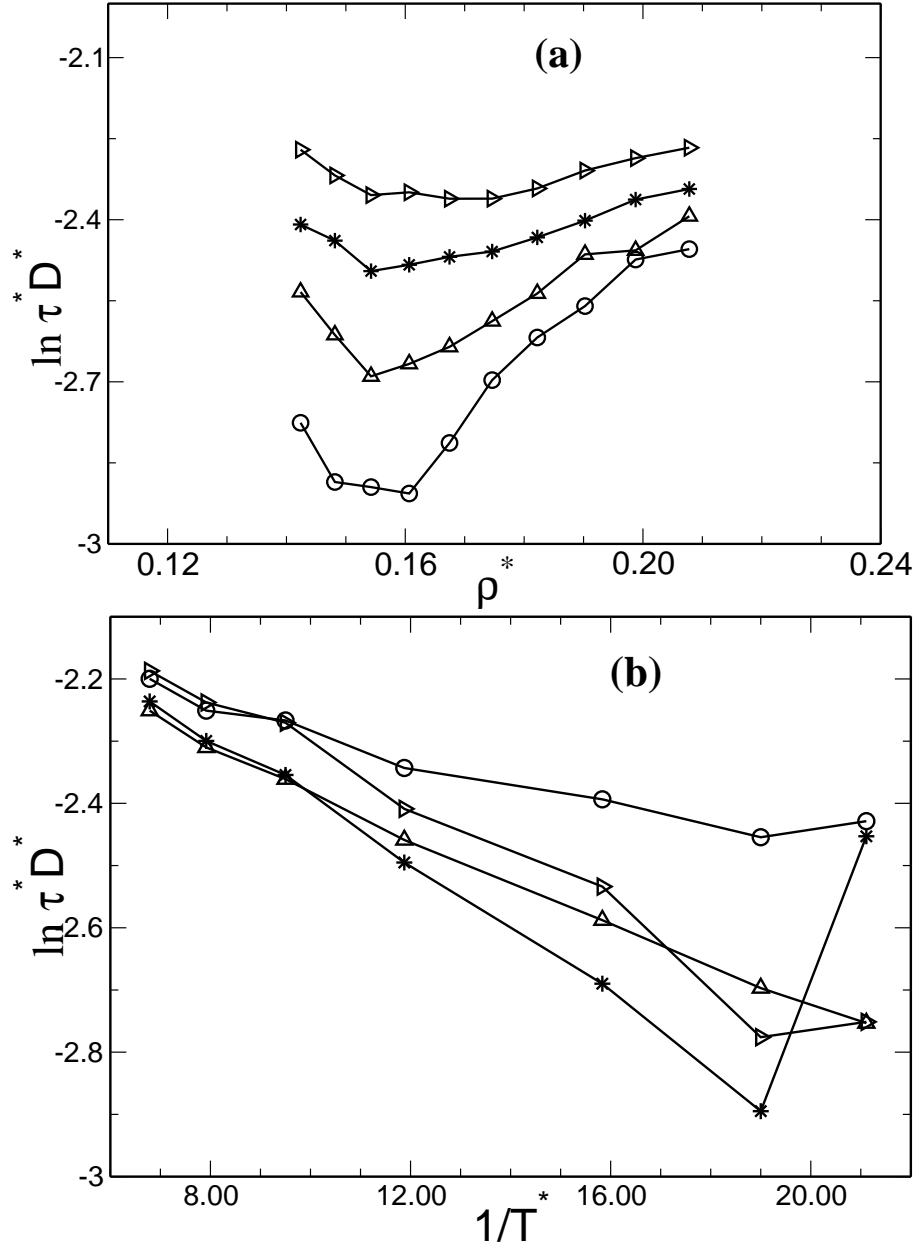

\begin{center}
\label{fig7ab}
\includegraphics[clip,width=12cm]{lnDtau-144.eps}
\includegraphics[clip,width=12cm]{lnDtau-1T-144.eps}
\end{center}
\caption{(a) Product $\tau_r^* D^*$ as a function of
density for four isotherms 
(from top to bottom $T^{*} = $ 0.105, 0.084, 0.063, 0.052)
 for $n=144$ steps (b) Product $\tau_r^* D^*$ as a function of the 
inverse of
temperature along isochores for $n=144$ steps (circles: $\rho$ = 0.208,
up triangles:  $\rho$ = 0.175, stars:   $\rho$ = 0.154,
right triangles:  $\rho$ = 0.142) }
\end{figure}

\begin{figure}
\begin{center}
\label{fig7}
\includegraphics[clip,width=12cm]{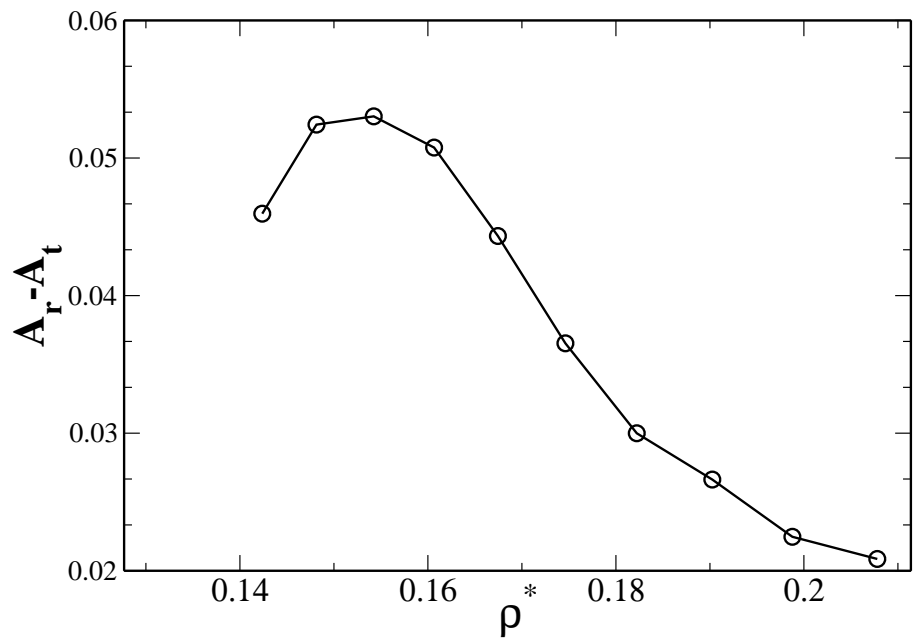}
\end{center}
\caption{Difference between the rotational and translational
activation energies, $A_r$ and $A_t$,
calculated  as a function of density from the product $\tau_r D$.}
\end{figure}


\begin{thebibliography}{40}
\bibitem{note2} The number of points included in the fit depends
on temperature: we took as cutoff for the exponential region
the largest time before the autocorrelation function become smaller
than its fluctuations.
\end{thebibliography}
\end{document}